# Evaluating Relayed and Switched Quantum Key Distribution (QKD) Network Architectures

Antonis Selentis, Nikolas Makris, Alkinoos Papageorgopoulos, Persefoni Konteli, Konstantinos Christodoulopoulos, George T. Kanellos and Dimitris Syvridis

***Abstract*— We evaluate the performance of two architectures for network-wide quantum key distribution (QKD): Relayed QKD, which relays keys over multi-link QKD paths for non-adjacent nodes, and Switched QKD, which uses optical switches to dynamically connect arbitrary QKD modules to form direct QKD links between them. An advantage of Switched QKD is that it distributes quantum keys end-to-end, whereas Relayed relies on trusted nodes. However, Switched depends on arbitrary matching of QKD modules. We first experimentally evaluate the performance of commercial DV-QKD modules; for each of three vendors we benchmark the performance in standard/matched module pairs and in unmatched pairs to emulate configurations in the Switched QKD network architecture. The analysis reveals that in some cases a notable variation in the generated secret key rate (SKR) between the matched and unmatched pairs is observed. Driven by these experimental findings, we conduct a comprehensive theoretical analysis that evaluates the network-wide performance of the two architectures. Our analysis is based on uniform ring networks, where we derive optimal key management configurations and analytical formulas for the achievable consumed SKR. We compare network performance under varying ring sizes, QKD link losses, QKD receivers' sensitivity and performance penalties of unmatched modules. Our findings indicate that Switched QKD performs better in dense rings (short distances, large node counts), while Relayed QKD is more effective in longer distances and large node counts. Moreover, we confirm that unmatched QKD modules penalties significantly impact the efficiency of Switched QKD architecture.**

## I. INTRODUCTION

QUANTUM Key Distribution (QKD) is among the most mature solutions to counter the threat posed by quantum computers to traditional network security, which could lead to data compromise [1]. Commercial deployments typically utilize Prepare and Measure QKD (PM-QKD) protocols to create QKD links and provide endpoints with quantum generated keys. In the beginning QKD networks relied on static point-to-point (P2P) configurations wherein quantum keys are generated in two quantum adjacent network nodes. However, deployed networks require quantum-level security between multiple communicating pairs. A baseline approach is to populate every pair of nodes with a dedicated QKD modules pair and a dedicated quantum fiber link to create a *fully connected* QKD network architecture. For a network of $N$ nodes and all-to-all secure communication, this approach requires $N{\cdot}(N\text{-}1)/2$ QKD module pairs. Thus, it becomes highly inefficient as the network becomes larger ($N$ increases), both in terms of complexity and cost.

Given the reach limitations of P2P QKD links [1], enabling key exchange over long distances requires one or more intermediate trusted nodes (ITU Y.3830 standard [3]). These nodes use back-to-back QKD (Alice-Bob or inverse) modules to forward keys via a key-relay function. More broadly, trusted nodes can relay keys not only to extend the reach of a single QKD link but also to support multiple relay paths for end-node pairs without a direct quantum fiber link. One such example is shown in Fig. 1a. A trusted and intelligent key management service (KMS) layer is required to manage these operations, with standardized protocols recently established for this purpose [4]. The KMS server at each node maintains buffers (referred to as Quantum Key Pools -QKPs) for the crossing relayed paths and implements the relay function (e.g., XORing keys from attached QKD links). The *Relayed* QKD architecture enables multi-nodes communication with relatively few QKD modules, requiring as many module pairs as the number of links. Thus, in a typical mesh quantum network with O($N$) fibers it requires O($N$) QKD modules. Relayed QKD can significantly extend the reach and secure more users cheaper, making it a candidate architecture for late QKD network deployments. The network of [5] employs 32 trusted relays to extend the network's coverage. In China, a 46-node network was recently demonstrated in a metropolitan area by utilizing a relatively small number (three) of interconnected trusted nodes, each serving as a hub in a star topology serving a subset of end-nodes [6].

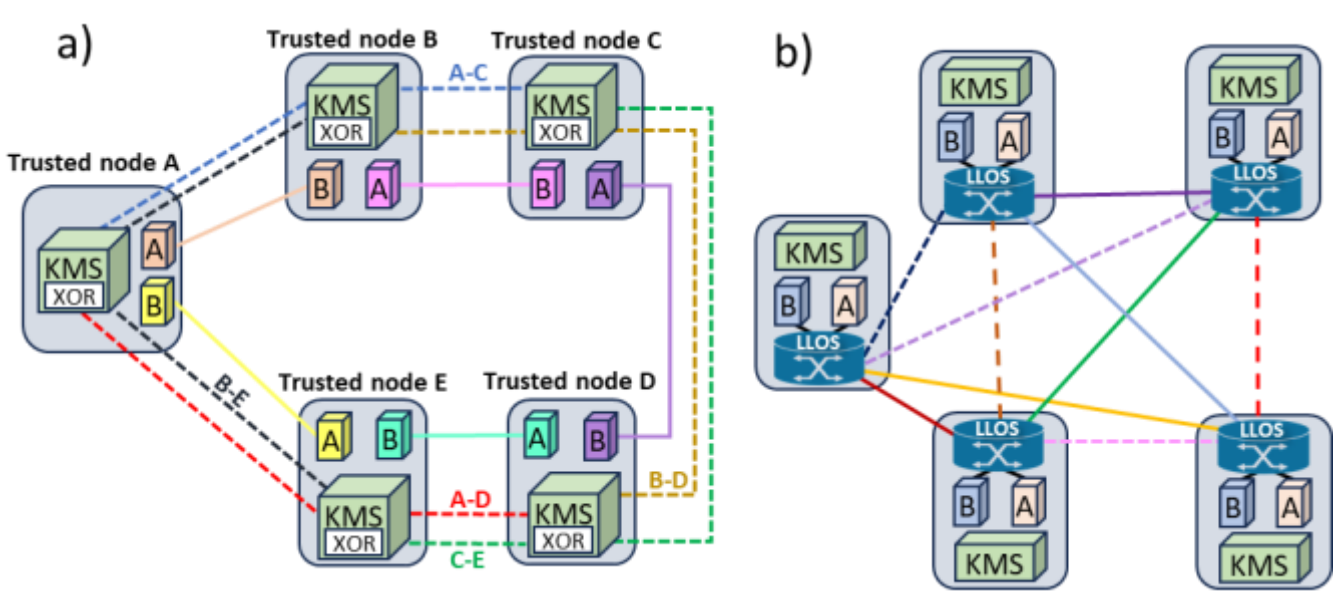

Fig. 1: a) Relayed QKD architecture in a 5-node network example, and b) Switched QKD architecture using 5 optical switches, one in every node.

While promising, Relayed QKD is inefficient in certain circumstances. When keys are relayed across a path, the relay function consumes keys of each QKD link comprising the path (with One-Time-Pad), reducing thus the achievable consumed SKR. This problem becomes significant for networks with a high number of nodes and extensive key relaying chains.

As an alternative, *Switched QKD architecture* uses optical switches (OS) at each node to enable direct optical connections/ QKD links between node pairs [7]. Fig. 1b shows a Switched QKD network example. As prepare and measure (PM)-QKD protocols only work in pairs, a QKD link is established between two nodes for a limited time, after which it is released, by reconfiguring the OSs, and the modules are repurposed to establish other QKD links. This approach allows a Time Division Multiplexing (TDM) provision of QKD links, effectively allowing $N$ QKD module pairs to be shared in different times and enable all-to-all key generation. In Switched QKD, the KMS servers at the active QKD link end-nodes store keys in Quantum Key Pools -QKPs during the active session and use them to maintain secure connectivity between the two nodes even when the QKD link is released. To ensure availability the QKP should be replenished by properly configuring the QKD links. Recently, several studies [8], [9], [10] considered the need for and proposed scheduling algorithms to efficiently manage the sessions of QKD links.

Compared to a fully connected QKD topology, Switched yields cost savings since it uses fewer QKD modules. Moreover, compared to the Relayed QKD architecture it offers the advantage of generating end-to-end quantum keys, since it does not rely on trusted nodes. However, its successful deployment requires meeting certain technical and operational requirements. Firstly, low loss optical switches (LLOS) need to be deployed, considering the exponential SKR generation decay as a function of attenuation [12]. Also, it is crucial to account for other impacts including the time needed for OS switching and QKD modules initialization procedures. Finally, QKD modules must support arbitrary communication, enabling any quantum transmitter (Alice) to link with any receiver (Bob). However, unmatched Alice-Bob pairings reduce SKR generation due to wavelength misalignment, filter tuning, phase mismatching, timing, or other protocol-specific factors. For example, in the BB84 protocol with decoy states and phase encoding [11], the transmitter's and receiver's interferometers require precise tuning which might not be satisfactory for an arbitrary pairing of modules. We demonstrate a high SKR generation deviation in the experimental part of this work.

Taking those limitations into account, we performed a theoretical analysis of both Relayed and Switched QKD architectures. We assumed a uniform ring network with equal length, SKR generation for adjacent nodes and opted for a fair (equal) allocation of keys for all-to-all demands. We derived optimal configurations and analytical formulas and used them to compare the two architectures and determine which is more suitable for different network scenarios. Our results indicate that the Switched architecture achieved significantly higher SKR consumption for small distances and large node counts. Conversely, the Relayed architecture achieved higher SKR consumption for large distances and medium to large node counts. Our findings also verified that unmatched QKD module penalties significantly degrade the efficiency of the Switched architecture. Thus, to leverage the advantages of Switched QKD, the industry should develop QKD modules capable of arbitrary connections with minimal performance penalties.

## II. Experimental Setup and Performance of Matched and Unmatched QKD Module Pairs

As mentioned earlier, pairing arbitrary QKD modules is essential in the Switched QKD architecture. However, typically, devices come in matched pairs, and thus pairing two unmatched QKD modules may result in a reduced SKR generation. To better understand this impact, we experimentally assessed the performance of matched and unmatched QKD module pairs of three different vendors employing different QKD technologies. Specifically, the tests were conducted on: i) Toshiba QKD 4.2-MU/MB employing phase encoding with decoy states (PDS) and a link budget up to 24 dB, ii) ID Quantique Clavis XG employing time-bin phase encoding (TBP) and a link budget up to 30 dB, and iii) Think Quantum Quky employing polarization-based encoding (PB) and a link budget up to 20 dB.

Initially we measured the matched pairs performance in terms of SKR generation. Fig. 2 shows the experimental measured SKR as a function of the link budget / attenuation, which was increased in steps of 0.5 dB, for the three vendors. Using these measurements, we applied linear regression to obtain SKR generation as a function of (continuous) attenuation $A$ (plotted in Fig. 2 with solid lines) We refer to such function, as the SKR generation function $G(A)$. We also plot the theoretical BB84 curve based on [12], that is implemented by all three of the QKD systems, but with Tx and Rx parameter set to approximate the Toshiba - PDS modules. This theoretical $G(A)$ function was used in the theoretical performance comparison of Switched and Relayed QKD architectures, presented in Section IV.

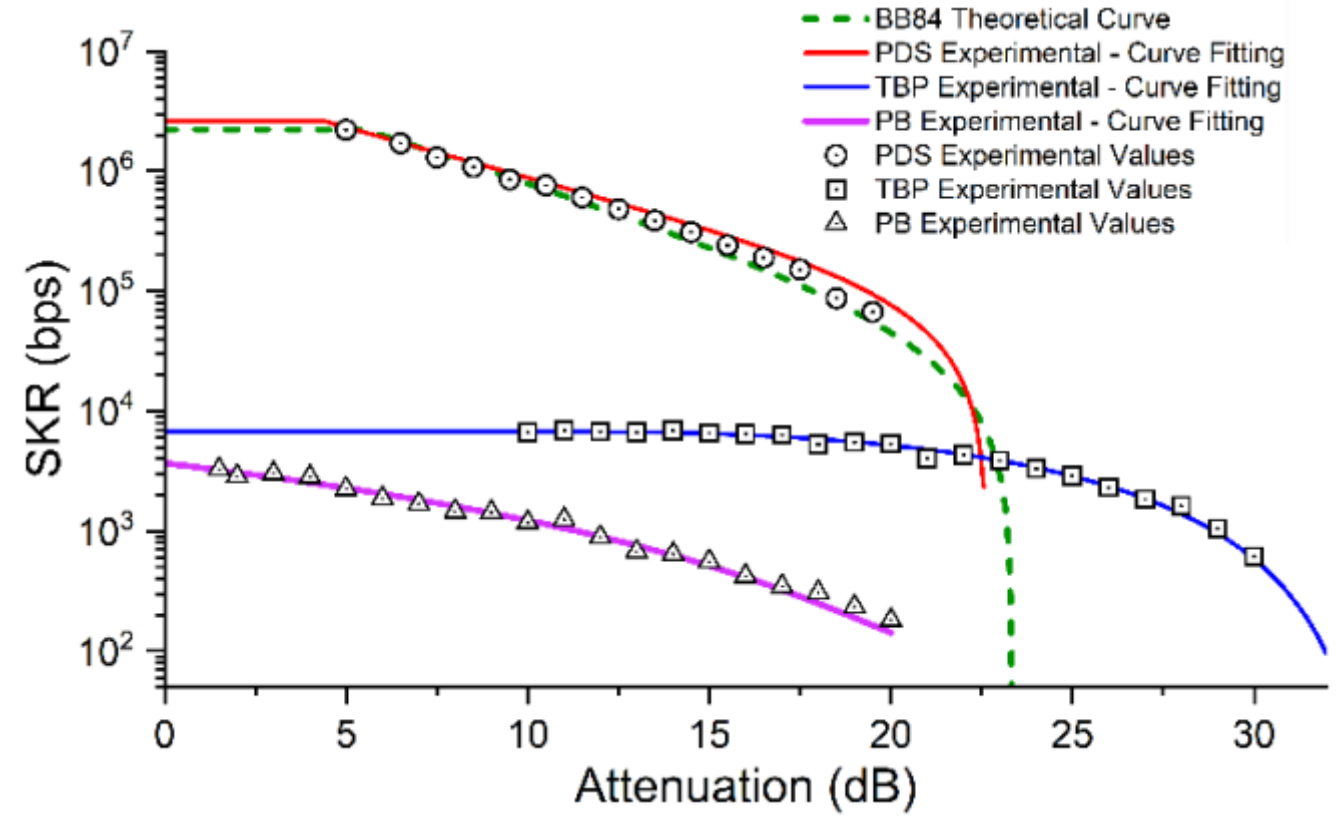

Fig. 2: SKR as a function of attenuation (in dB) for various QKD systems. The experimentally measured values for three commercial modules are plotted using corresponding symbols, while solid lines are their fitted curves. The green dashed line represents the theoretical SKR curve for BB84 with decoy states and Tx and Rx parameters based on PDS modules.

QKD modules, like most communication equipment, are designed to operate in a specific regime, most times optimized for long-distance operation. The QKD modules used in our experiments operate effectively at a receiver dynamic range. To be more specific, the Toshiba QKD 4.2-MU/MB with phase encoding with decoy states (PDS) operate between -6 dBm to -22 dBm. If the received power is higher, attenuators should be added to bring it below -6 dBm, otherwise we can damage the Rx. In other words, at short distances/low losses, the SKR generation experiences limitations due to detector saturation [16]-[18]. Additionally other factors such as the implemented error correction protocol and the processing power of the GPU or CPU used for error correction can also contribute to the SKR generation limitation in low losses regime [19]. So, most commercial modules require an optical attenuator to ensure a minimum operational power level. According to the datasheets of the vendors we tested: PDS modules require 6 dB attenuation as discussed above, TBP modules require 10 dB attenuation for the quantum, while PB modules have no attenuation requirement. These specifications were verified in our lab and are observed in Fig. 2 with the flat SKR generation curves for the TBP and PDS from 0 dB up to the corresponding attenuation value. While the flatness of SKR generation at low losses may seem disadvantageous, it actually becomes beneficial for the Switched as opposed to the Relayed architecture, as will be discussed in our comparison study.

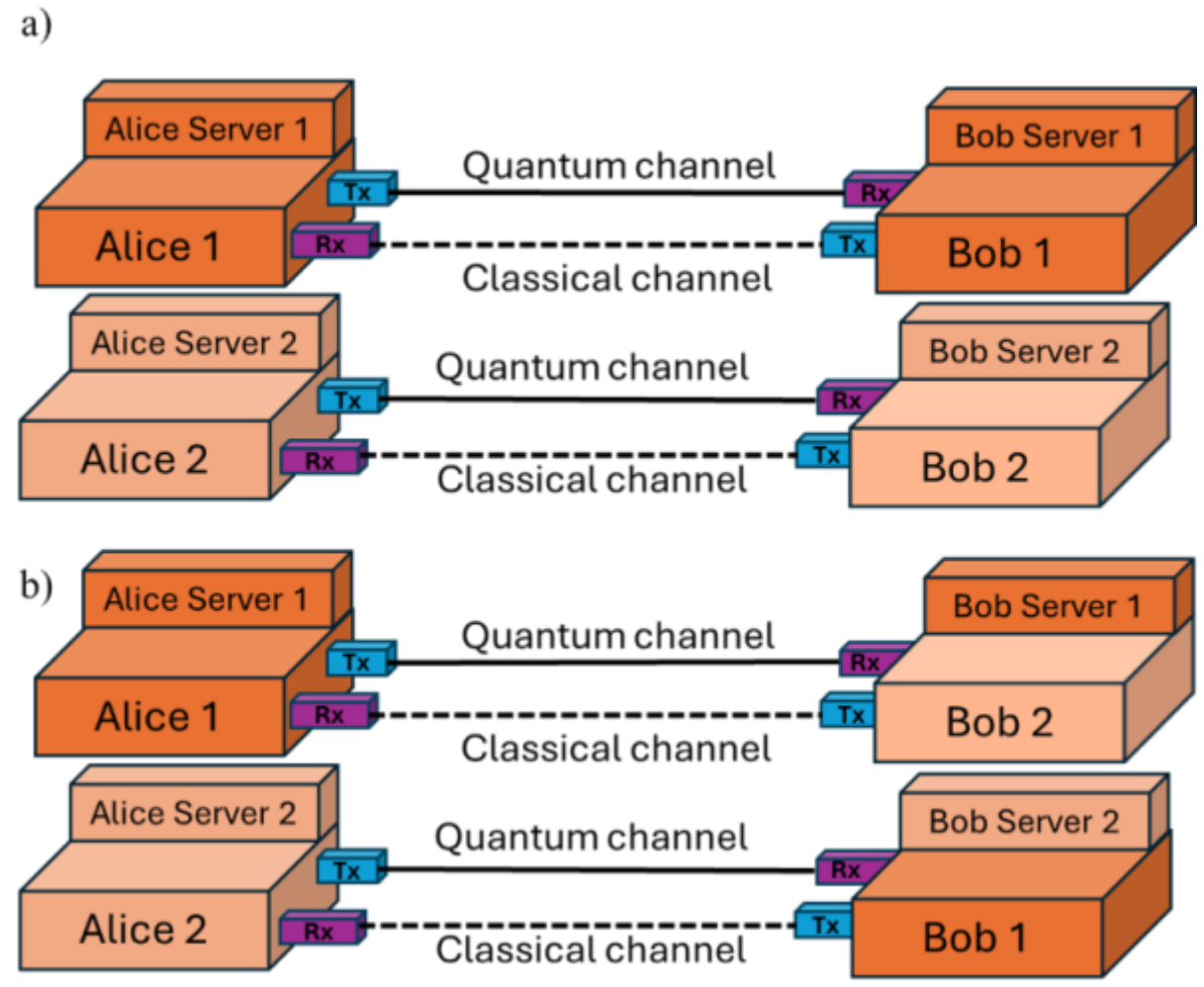

Fig. 3: a) Toshiba matched QKD pairs configuration (bar configuration) vs b) unmatched QKD pairs configuration (cross configuration for Switched QKD).

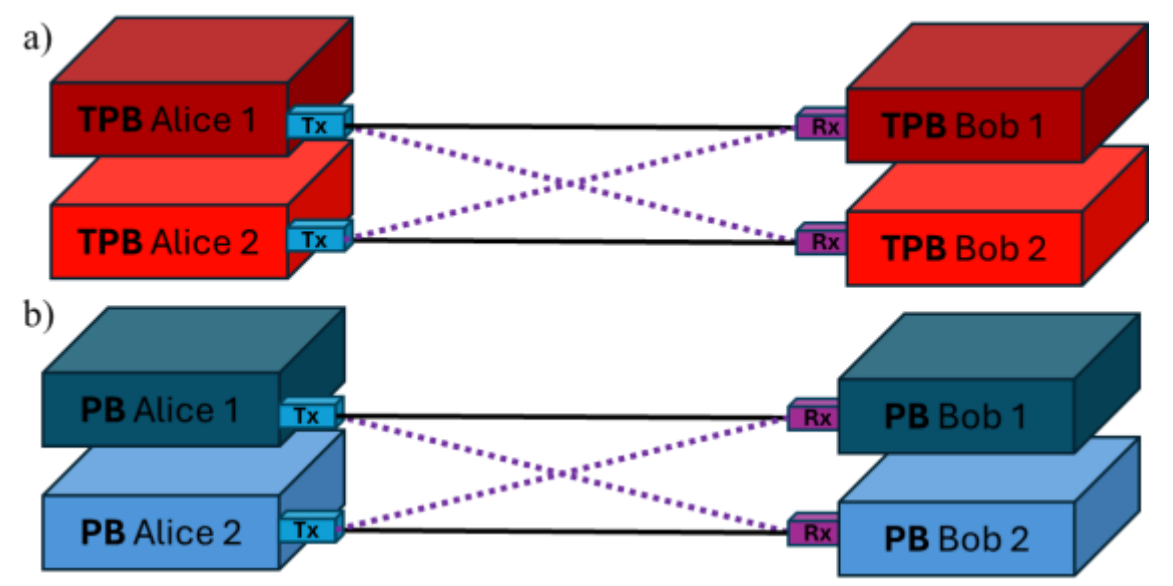

Fig. 4: Matched QKD pairs configuration (bar configuration) is illustrated with solid line while unmatched QKD pairs configuration (Switched or 'cross' configuration) with dashed line for a) TPB, b) PB modules.

Next, we tested both matched and unmatched (Switched) pairings at specific attenuation levels for the three vendors.

For the PDS system, we first tested the *bar* configuration where two quantum transmitters (Alices) were connected to their matched receivers (Bobs). Essentially, *bar* configuration refers to two matched pairs. So, for the *bar* configuration Alice 1 was connected to its matched pair Bob 1 (A1B1) via a 6.4 km link along with 6 dB additional attenuation to emulate additional fiber length and LLOS. The second pair Alice 2 was connected to Bob 2 (A2B2) via an 8 km link plus a 6 dB attenuation. In each pair, the QKD quantum channel was emitted at 1310 nm and coexisted with the classical and service channels at 1529 nm and 1530 nm, respectively. PDS QKD pairs also included a back propagation classical channel with direction from Bob to Alice transmitted at 1528 nm.

Then, for the Switched QKD case operation we changed the connectivity to the *cross* configuration. We established QKD connections between Alice 1 to Bob 2 and between Alice 2 to Bob 1. Thus, cross configuration refers to two unmatched pairs. The quantum links distance was kept the same for the Alices to make a direct comparison between the two cases. An important note is that in our experimental testbed we manually swapped only the optical QKD engines and maintained the original servers' connectivity, as shown in Fig. 3. This was done because the servers were locked, for security purposes, to only communicate in matched pairs. Such configuration could have some effect on the measured SKR performance. The implemented phase encoding protocol necessitates precise alignment and perfect matching between the differential interferometers (DI). Typically, adjustments to the DIs are carried out during initialization by the servers. The servers may rely on stored data for their corresponding optical engines that if swapped could cause some misalignment issues.

TABLE I

AVERAGE VALUES FOR GENERATED SKR AND QBER FOR MATCHED (BAR) AND UNMATCHED (CROSS) PAIRS

| Metric | Bar Configuration | | Cross Configuration | |
|---|---|---|---|---|
| | A1-B1 | A2-B2 | A1-B2 | A2-B1 |
| MU (Toshiba) – PDS (Attenuation: 6 dB) | | | | |
| SKR (bps) | 481 k | 722 k | 3.3 k | 27 k |
| QBER (%) | 3.13 | 2.93 | 6.7 | 5.7 |
| SKR S.D. (bps) | 34 k | 124 k | 7.6 k | 11.2 k |
| QBER S.D. (%) | 0.002 | 0.008 | 0.139 | 0.111 |
| SKR Δ (%) | | | 99.31 % | 96.26 % |
| Clavis XG (IDQ) – TBP (Attenuation: 10 dB) | | | | |
| SKR (bps) | 6264 | 6628 | 6085 | 6999 |
| QBER (%) | 0.73 | 0.69 | 0.75 | 0.51 |
| SKR S.D. (bps) | 208.90 | 208.88 | 199.26 | 221.62 |
| QBER S.D. (%) | 0.083 | 0.081 | 0.086 | 0.069 |
| SKR Δ (%) | | | 2.86 % | -5.30 % |
| Quky (TQ) – PB (Attenuation: 0 dB) | | | | |
| SKR (bps) | 3371 | 6639 | 4385 | 5639 |
| QBER X (%) | 0.82 | 0.80 | 0.90 | 0.44 |
| QBER Z (%) | 1.80 | 0.77 | 1.57 | 0.95 |
| SKR S.D. (bps) | 402 | 335 | 730 | 524 |

| QBER X S.D.(%) | 0.42 | 0.31 | 0.47 | 0.22 |
|---|---|---|---|---|
| QBER Z S.D.(%) | 0.56 | 0.21 | 0.55 | 0.69 |
| SKR Δ (%) | | | -23.12 % | 15.06 % |

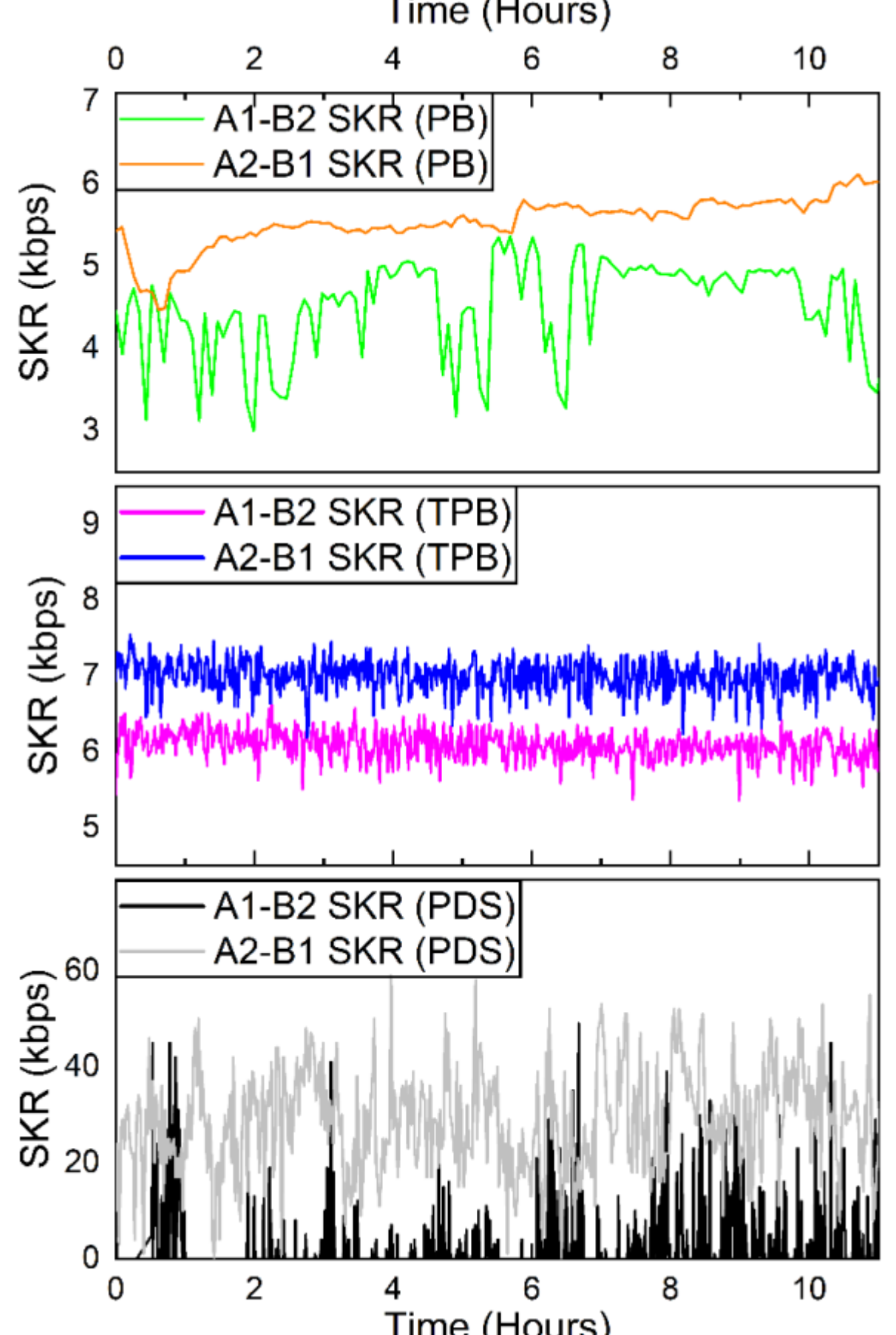

Fig. 5 SKR performance of the unmatched A1-B2 and A2-B1 pairs (cross configuration) for (a) PB, (b) TPB and (c) PDS modules as a function of time for 12-hours interval operation.

After measuring the performance of the QKD modules for approximately 11 hours, we averaged the QBER and SKR generation values and present them in TABLE I. The results show varying performance deviations between the matched (bar) and unmatched (cross) pairing scenarios for the three tested QKD vendors. Fig. 5 shows the SKR generation evolution over time for the unmatched (cross) pairings for the three QKD vendors.

For the PDS modules, the measured average SKR and QBER reveal nearly a two-order-of-magnitude deviation between the matched and unmatched pairs (see SKR Δ in Table I). These measurements refer to the specific modules that we tested in our lab, which were manufactured to operate only in matched pairs, without considering their potential use in a Switched network. The PDS modules in the *cross* configuration operated at their limits, as indicated by the increased standard deviation (Table I) and the multiple zero SKR values observed for the PDS A1-B2 pair in Fig. 5 (c). On the other hand, the TPB and PB modules, which are designed to operate in unmatched pairs, exhibit smaller but still non-negligible deviations. Specifically, for the PB pair, the SKR performance deviation (SKR Δ) was +15% and -23% for the two tested pairs, respectively. Although less severe compared to the PDS case, and even negative in one unmatched pair, these deviations remain relatively high and could influence the key consumption performance. In contrast, for the TPB pair, the SKR deviation (SKR Δ) was almost negligible, +2.86% and -5.30% for the two respective pairs. Thus, our findings underscore the significance of considering the varying SKR generation that stems from using unmatched pairs in Switched QKD networks.

## III. Comparing Switched to Relayed Architectures

The observed variations in SKR generation and the differing behavior across QKD vendors when unmatched pairs are connected, as reported in the previous section, highlight the need for further investigation. To this end, we conducted a numerical analysis comparing the Switched to the Relayed QKD network architectures.

We made specific assumptions that simplify the network-wide key distribution, enabling us to construct optimal configurations and derive analytical formulas to directly compare the two architectures. We assumed a uniform ring network, with equal distances and attenuation between adjacent nodes, uniform SKR generation, and all-to-all key consumption demands. Although real networks are heterogeneous with non-uniform SKR generations, our comparison provides a directional guide and a first-order approximation for more realistic scenarios.

We considered a 3-layer network. The first layer is the quantum layer, which includes the quantum modules and dedicated quantum fiber links. Two nodes, $(n_i, n_j)$ form a QKD link through the communication of an Alice QKD module at node $n_i$ and a Bob at node $n_j$. While the link is active, symmetric quantum keys are generated at the endpoints. The second layer is the key management system (KMS). There is a KMS server at each node that takes the keys generated from the local QKD modules and stores them in quantum key pools (QKPs) and/or transfer encrypted keys to another KMS server as an intermediate step of creating key relay paths (depending on the network configuration). Then, at the third layer, secured application entities (SAEs) use the symmetric keys from the KMS servers to encrypt their communications over conventional channels. The orchestration is done by the KMS, which maintains separate Quantum Key Pools (QKPs) for every SAE pair. Note that we use the term *generation* to refer to the QKD modules communication, the output of the QKD/quantum layer to the KMS servers. Conversely, we use the term *consumption* to refer to the use of keys by the communicating SAEs, so the output of the KMS layer to the SAEs.

We considered a ring network topology of $N$ nodes (with $N$ being odd to simplify the formulas), commonly used in metro networks, with equal adjacent nodes link lengths $L^e$, and equal attenuations $A^e=a_c \cdot L^e$, where $a_c$ is the attenuation coefficient. In both Switched and Relayed QKD network architectures, we assumed that each node is equipped with one Alice and one Bob module. The SKR generation between an Alice and Bob pair is given by a function, $G(A)$, of the attenuation $A$, as discussed in Section II – Fig. 2. We assumed the same SKR generation function, $G_{ij}=G(A_{ij})$, for each quantum communicating pair ($n_i$,$n_j$) where $A_{ij}$ is the attenuation of the pair at hand. We also assumed additional optical penalties $O$ for the Switched QKD to account for the additional losses due to optical switches. In the Switched, we also introduced a SKR generation penalty $P$ to capture the effect of arbitrary pairing of Alices and Bobs.

We also assumed an all-to-all nodes key consumption pattern, where, in a ring of $N$ nodes, each node has $N$-1 SAEs, with each SAE communicating with a SAE in every other node. Thus, for each pair of nodes ($n_s$, $n_d$), there is a corresponding SAE pair consuming keys. The KMS servers at nodes $n_s$ and $n_d$ maintain dedicated buffers (QKPs) with symmetric keys for ($n_s$,$n_d$) communicating SAEs. These keys are provided to SAEs at a *consumption SKR* $C_{sd}$. We assumed that we opt to maximize the minimum consumption rate, that is, max($\min_{sd}(C_{sd})$), across all $N(N$-1)/2 SAE pairs, ensuring fairness and enhanced overall network security. The QKD and KMS layers operate differently in the Switched and Relayed QKD networks, which are discussed individually in the following subsections.

### *A. Relayed QKD Network performance*

In the Relayed architecture, for a ring network, we assumed that the Alice of each node $n_i$ is connected to the Bob of next node $n_{i+1}$, and so on, creating a chain across the ring using $N$ module pairs. Each node has QKD links that generate keys exclusively with its (two) neighboring nodes at rates ($G_{i,i+1}$ and $G_{i,i-1}$, for the next and previous node) depending on the attenuation. According to our assumptions, all these QKD links have the same distance $L^e$ and attenuation $A^e$=$a_c \cdot L^e$, where $a_c$ is the attenuation coefficient and thus, the same SKR generation $G_{i,i+1}$=$G(A^e)$, for all $i$. The distribution of symmetric keys to remote (non-adjacent) nodes is facilitated through relaying at the intermediate nodes KMS servers, that act as trusted nodes. Specifically, suppose we aim to generate a quantum key for SAE pair ($n_i$,$n_{(i+k)\%N}$). Note that '%' is the modulo operation. All additions and subtractions are done in modulo $N$ arithmetic and will be suppressed in the following to simplify the notation. The KMS server of node $n_i$ generates a key and encrypts it (with One-Time-Pad -OTP e.g. by XORing) with a symmetric quantum key generated by QKD link ($n_i$,$n_{i+1}$) and sends it to the KMS server of node $n_{i+1}$ through a conventional channel. Then, in turn, the KMS server or $n_{i+1}$ decrypts using the quantum key of QKD link ($n_i$, $n_{i+1}$), obtains the initial key and encrypts it with a quantum key of QKD link ($n_{i+1}$,$n_{i+2}$), sends it to $n_{i+2}$, and so on. Eventually the key reaches $n_{i+k}$ and can be used for a quantum-safe communication for the SAE pair of $n_i$ and $n_{i+k}$. Note that we assumed that intermediate nodes relay keys with OTP e.g. XOR, rather than encrypting the keys with a cipher of a specific key length (e.g., AES). This ensures that the KMS provides quantum-secure keys end-to-end, assuming the trust to intermediate nodes.

To perform the relay, the KMS selects one or more paths for relaying keys between two non-adjacent quantum nodes. Relaying keys over a path using OTP consumes keys from all the QKD links along the path. Assuming one path $p_{sd}$ connecting ($n_s$,$n_d$), the total end-to-end key consumption rate $C_{sd}$ is the minimum of the intermediate QKD links contributions: $C_{sd} = \min_{i,j \in p_{sd}} C_{ij}^{sd}$ where $C_{ij}^{sd}$ is the contribution of QKD link ($n_i$,$n_j$) on the relay path $p_{sd}$. When multiple paths are used between $s$ and $d$, we add their contributions. The KMS decides for all SAE pairs with non-direct QKD links, the relay paths and the key contributions of the QKD links to each relay path, accounting for the generation rate of each QKD link. This optimization problem is a variation of the multicommodity flow problem, a typical problem in graphs and networks [20], [21]. Fig. 6 shows an example of the KMS system depicting QKPs and relay function for a simple 3 nodes network.

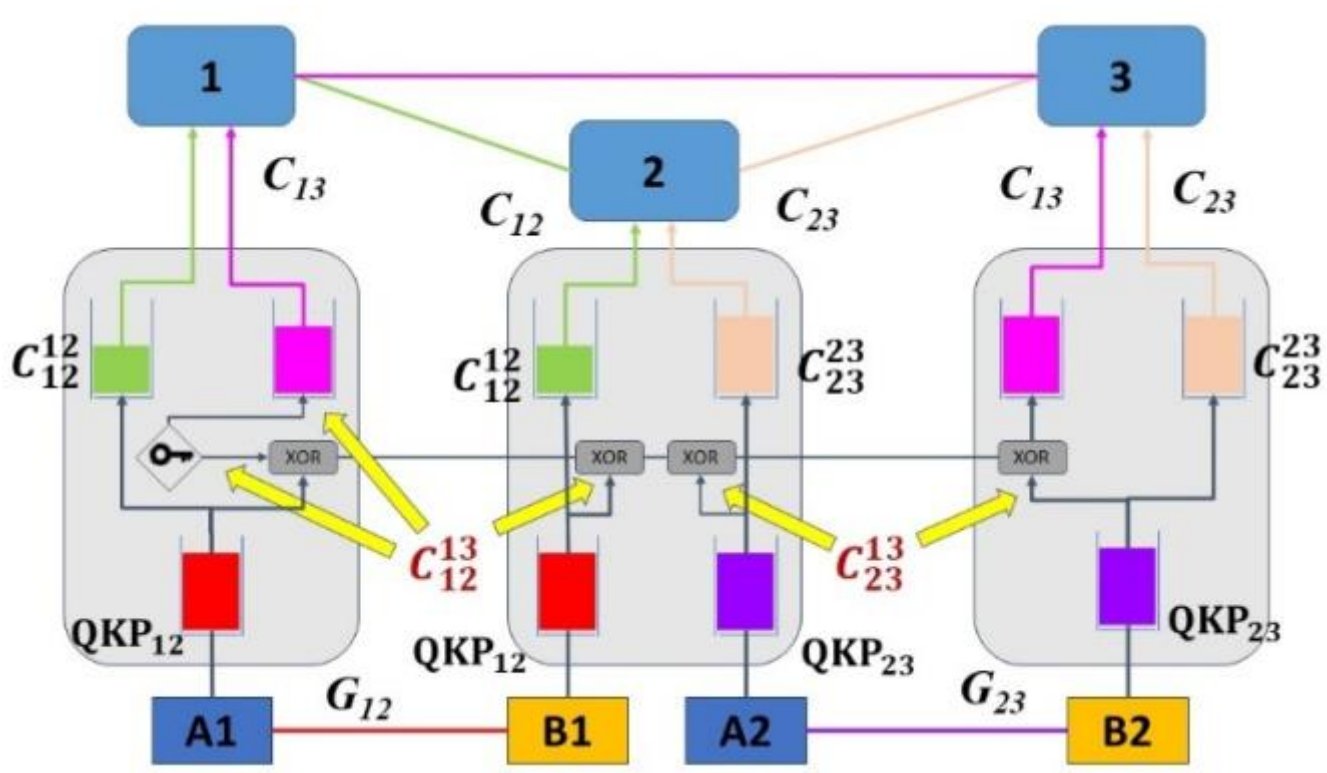

Fig. 6: Details of the Relayed QKD architecture for 3 nodes with only two QKD module pairs (Alice A1 matched with Bob B1, and Alice A2 matched with Bob B2), supporting all-to-all secure communication (3 SAE pairs). The KMS implements two pooling levels. The first level is fed with keys generated at a rate $G_{ij}$. These keys are then split into contributions for each relay path using the QKD link ($n_i$,$n_j$). The end-to-end consumption rate $C_{sd}$ of the relay path $p_{sd}$ is determined by the minimum of contributions of the QKD links along the path.

In a ring topology, and considering our assumptions, these calculations are simplified. We use only the shortest path: for each node $n_s$, the remote node $n_{s\pm k}$, with $1 \le k \le N$-1/2, is $k$ hops away, since SKR generation is bidirectional. Also, since all QKD links have the same generation rate and we opt to maximize the minimum consumption rate, we can equalize the consumption rates for all SAE pairs. To do so, we equally divide the keys of each QKD link to all relaying paths crossing that QKD link, that is we equalize all $C_{ij}^{sd}$. To be more specific, the keys generated by QKD link ($n_i$,$n_{i+1}$) will be used by $n_i$ to communicate with subsequent nodes $n_{i+1}$, $n_{i+2}$, and so forth, up to $n_{i+(N-1)/2}$, and thus for ($N$−1)/2 SAE pairs. Simultaneously, these ($n_i$,$n_{i+1}$) keys will be also used by the preceding nodes, and in particular, by node $n_{i-1}$ for communicating with ($N$-1)/2-1 other nodes that are numbered higher than $n_i$, by node $n_{i-2}$ for communicating with ($N$-1)/2-2 other nodes that are numbered higher than $n_i$, and so on, in progressively fewer pairs. In total, QKD link ($n_i$,$n_{i+1}$) keys will feed $\sum_{j=0}^{\frac{N-1}{2}-1}(\frac{N-1}{2} - j) = \frac{N^2-1}{8}$ different SAE pairs. Therefore, considering that $G(A^e)$ represents the generation rate between two neighboring nodes, we divide that equally to all relaying paths to achieve our objective, and thus the consumption rate of the Relayed QKD architecture will be

$$C_R = 8 \cdot \frac{G(A^e)}{(N^2-1)},\ A^e = a_c \cdot L^e, \tag{1}$$

*B. Switched QKD network performance*

In Switched QKD, for the quantum layer we assumed separate and direct fibers from each node to any other node, and thus the quantum network is a ring with the addition of fibers over the chords. Note that similar analyses can be done under different assumptions, e.g. using multiple fibers in cables around the ring or by using a single fiber while avoiding simultaneous key sessions on common links, etc. Each node is equipped with an optical switch connected to the $N$-1 (chord) fibers to all other nodes. The optical switch at a node can be configured to connect the Alice/Bob module of the node to a selected fiber. At the other end, the corresponding switch is configured to connect that fiber to its Bob/Alice module, thereby establishing the QKD link. We assumed that each node can simultaneously utilize two fibers—one for its Alice and one for its Bob. We targeted to operate the Switched QKD network in a periodic manner with a period $T$. When configuring a QKD link ($n_i$,$n_j$), key material is generated at a rate $G_{ij}$, extracted, and stored in the respective QKPs at both nodes. Although keys are produced for a specific duration $D_{ij}$ within a period $T$, they are continuously consumed at a rate $C_{ij}$. To prevent key starvation, the generated keys during $D_{ij}$ must suffice for the entire period $T$ ($G_{ij}\cdot D_{ij}$<=$C_{ij}\cdot T$).

Fig. 7 illustrates a 3-node network. In the first time instant (Fig. 7a), nodes 1 and 2 (A1-B1 – matched pair) and nodes 2 and 3 (A2-B2 – matched pair) generate keys. In the second time instant (Fig. 7b) nodes 1 and 3 (A1-B2 – unmatched pair) generate keys. The keys are stored in the corresponding QKPs as shown and continuously consumed at (symmetric) rates $C_{12}$, $C_{13}$, and $C_{23}$. In the general case, a scheduling algorithm is required to determine the appropriate switching times, avoid key starvation, and maximize the SKR consumption. However, in a uniform ring network with uniform SKR generation, scheduling becomes straightforward, as discussed below.

We will construct a periodic schedule for the Switched QKD network with the goal of achieving equal and maximum consumption rates among all SAE pairs. For each node $n_i$, its Alice module is scheduled to establish QKD links with nodes from $n_{i+1}$ to $n_{(i+N/2)}$, while its Bob module is schedule to connect with nodes from $n_{i-1}$ to $n_{(i-N/2)}$.

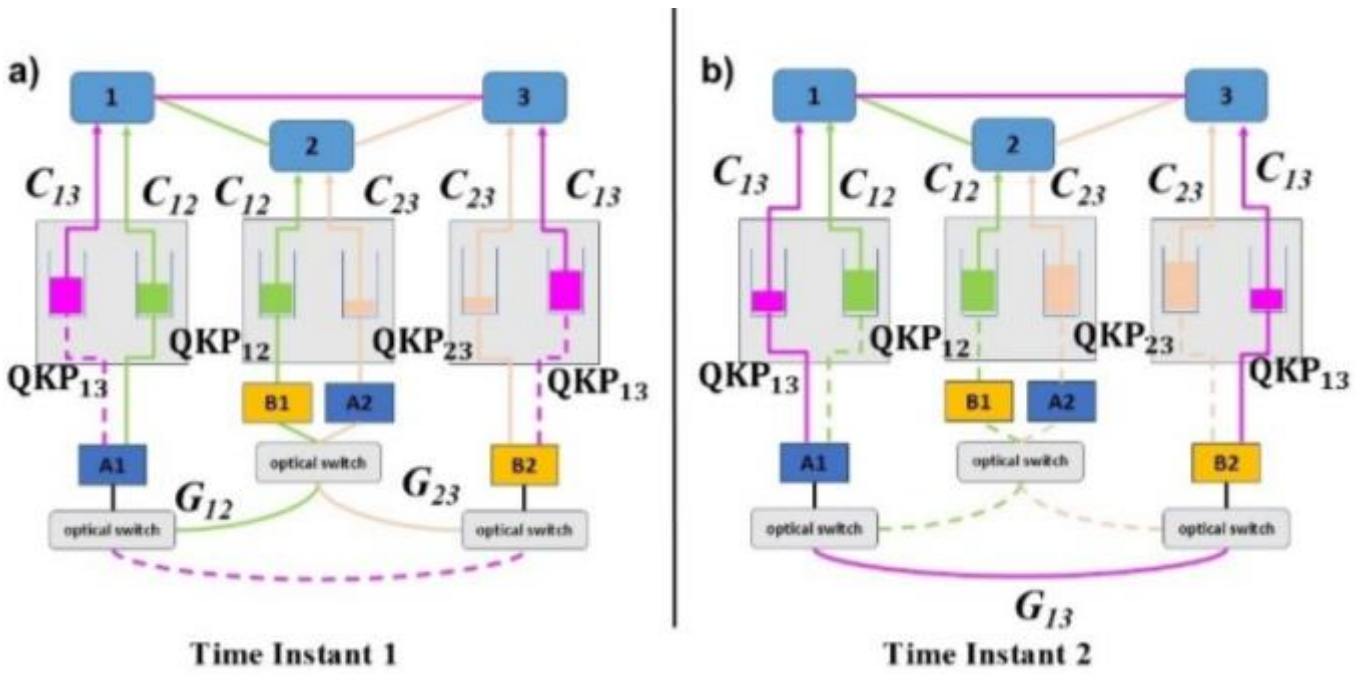

Fig. 7: Details of the Switched QKD architecture for 3 nodes with only two QKD module pairs, supporting all-to-all secure communication (3 SAE pairs). The KMS servers implement quantum key pools (QKPs) for all pairs, fed with keys at a rate $G_{ij}$ only when direct QKD links are established. These keys can be consumed even after the links are released, for example, at a constant consumption rate $C_{ij}$. Two-time instants with different sets of QKD links are shown, demonstrating how all QKPs are replenished.

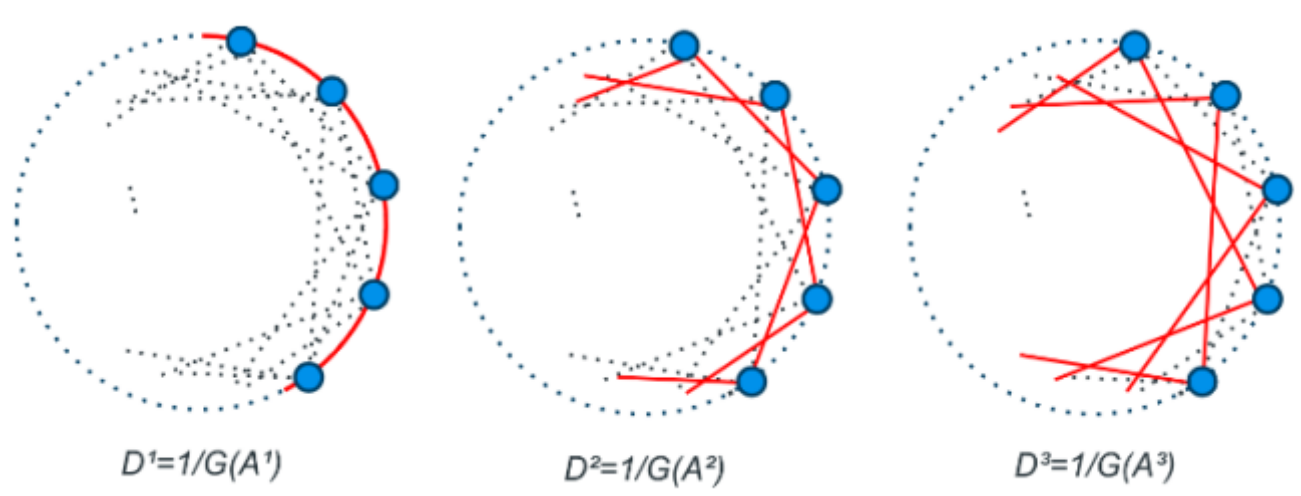

Fig. 8: Scheduling phases for the Switched QKD network: establishing simultaneously all 1-hop distance, 2-hop distance, etc., QKD links.

Key generation sessions between all node pairs of the same $k$ hop-distance take place simultaneously, in the respective *k-hop phase*, as explained next. Given the uniformity, the attenuation of nodes that are $k$-hops apart in the ring, that is, of nodes ($n_i$,$n_{i+k}$) for every $n_i$ and 1≤$k$≤($N$-1)/2, are equal across the network. We denote that attenuation by $A^k$= $a_c\cdot$ *chord(k,*$L^e$*)* where *chord*() represents the distance (in km) between two nodes that are $k$ hops apart, measured along the chord of the ring network with adjacent node link length $L^e$. Thus, the generation rate is given by $G_{i,i+k}$=$G(A^k)$, for all $n_i$ and 1≤$k$≤($N$-1)/2.

We construct our schedule so that the QKD links between all $k$-hops distance nodes are established simultaneously, in the $k$-hops phase, with the same starting times and for duration $D_{i,i+k} = D^k$, for every $i$, and $1 \leq k \leq (N\text{-}1)/2$. Note that each node $n_i$ is included twice for a specific $k$-hop distance, as it participates in two pairs: ($n_i$,$n_{i+k}$) and ($n_{i-k}$,$n_i$), but both QKD links are established simultaneously, using Alice and Bob of node $n_i$, respectively. The schedule is constructed by starting from the $k$ = 1 phase, where all 1-hop QKD links are established for duration $D^1$, and then increase $k$ step by step up to ($N$-1)/2, to cover all node pairs. This is visualized in Fig. 8. After each $k$-hop phase, to move to the ($k$+1)-hop phase, a reconfiguration time $R$ is required. The total duration $T$ for generating and distributing keys across all node pairs is the sum of the durations $D^k$. Considering reconfiguration penalties, this is given by $T = \frac{R\cdot(N-1)}{2} + \sum_{k=1}^{(N-1)/2} \frac{1}{G(A^k)}$. The consumption rate $C_{ij}$ for nodes ($n_i$,$n_j$) should be lower or equal to the generated, $C_{ij} \leq D_{ij} \cdot G(A_{ij})/\ T$. Under our assumptions, all $k$-hops distant nodes have the same attenuation $A^k$ and SKR generation $G(A^k)$. Since they are served simultaneously and for the same duration $D^k$, they have the same consumption rate $C^k = D^k \cdot G(A^k)/\ T$.

To maximize the minimum consumption rate, we should equalize the consumption rates $C^k$ for all $1 \leq k \leq (N\text{-}1)/2$. Thus, we set $D^k$ = 1 bit/ $G(A^k)$ and obtain $C^k$ = 1 bit/ $T$. The intuition behind this is that distant hops, that have high attenuation, have low SKR generation and thus are given inversely proportional longer time. Therefore, each SAE pair and thus the consumption rate of the Switched QKD network will have a consumption rate of

$$C_S = \frac{1\text{ bit}}{\frac{R(N-1)}{2}+\sum_{k=1}^{(N-1)/2}\frac{1}{G(A^k)}} \tag{2}$$

This construction defines the relative durations of the different QKD links and enables equivalent optimal schedules by proportionally scaling all durations, with the period scaling accordingly while accounting for the reconfiguration times.

As discussed, in the Switched QKD architecture we add a penalty $O$ (dB) for the use of optical switches in the attenuation

$A^k$. Moreover, we have an additional performance penalty $P$ (dB), applied to the output of the SKR generation function $G(A)$ for the arbitrary matching of QKD modules. Thus, to be precise, in Eq. (2), $G(A^k)$ is replaced by $G(A^k+O)\cdot 10^{-P/10}$.

Recent studies proposed optimal integer linear programming (ILP)-based algorithms to design the Switched [8] and Relayed QKD networks [20] for generic network scenarios. The above analytical formulas were obtained under specific assumptions (uniform rings and uniform SKR generation function). We verified that for such assumptions the optimal ILP formulations of [8], [20], when adapted to optimize the minimum consumption rate, yield the same optimal consumption rates as our analytical formulas outlined above.

## IV. RESULTS

Using the analytical formulas introduced above we compared the Relayed and Switched QKD architectures. We assumed uniform rings with varying number of nodes ($N$=5 to 25) and varying adjacent node distances ($L^e$=1 to 35 km), with an attenuation coefficient ($a_c$) equal to 0.24 dB/km. As previously mentioned, the attenuation, used in the SKR generation function $G(A)$, is calculated between adjacent nodes in the Relayed architecture, so is equal to $A^e=a_c\cdot L^e$, for all links, and over the chord for $k$-distant hops in the Switched architecture, so is equal to $A^k=a_c\cdot chord(k,L^e)$. We used the theoretical $G(A)$ function with parameters driven by the Toshiba PDS QKD modules. Driven by the PDS modules, the quantum transmitters were assumed to operate at 0 dBm and receivers were assumed to have maximum acceptable received power of $R_A$=-6 dBm, and thus a constant SKR generation for distances below 25 km, what we call *flat region K,* as discussed in Section II. Moreover, as previously discussed, for the Switched network, we assumed an extra attenuation $O=2$ dB due to low-loss optical switches and a performance penalty $P=2$ dB due to QKD module mismatch (Section II). We also assumed a reconfiguration time of $R$=5 minutes and a normalized period of $T$= 180 minutes.

We compared the two architectures by calculating the maximum minimum SKR consumption $C_R$ for the Relayed and $C_S$ for the Switched, based on Equations (1) and (2) respectively, for every ring scenario with specific number of nodes $N$ and adjacent length $L^e$. We focus the presentation of the results on the relative performance of the two architectures and to visualize this, we plotted the relative consumption rate gain ($f$) of Switched over Relayed QKD,

$$f=100\cdot (C_S-C_R)/\max(C_S,C_R) \quad (3)$$

ranging from 100 (where the Switched outperforms the Relayed) to -100 (opposite).

Fig. 9 shows the relative performance of the Switched and Relayed architecture for indicative ring scenarios ($N$=5 to 25 nodes, and $L^e$ = 1 to 35 km distance). As evident, Switched QKD outperforms Relayed QKD in every ring scenario with small distances between nodes. For $N$=5 nodes and all $L^e$ the improvement remains stable due to the SKR generation function flat region $K$ (=25km). For the rest, the Switched architecture performs notably better ($f$>25), for $L^e$ ≤5 km distance and all examined $N$, and for $L^e$ ≤7.5 km distance and $N$ <= 13 nodes. The improvement becomes significant ($f$>50), in large scale networks ($N$≥17) with very small node distances ($L^e$<5). On the other hand, Relayed QKD outperforms ($f$<-50) Switched QKD in scenarios of relatively small networks ($N$=9) and $L^e$>20km, medium sized networks ($N$=13, 17) and $L^e$>15km, and larger networks ($N$≥21) and $L^e$>10km. Moreover, Relayed achieves a relatively better performance ($f$<-25) in scenarios of $N$=5 nodes and $L^e$=25 km, $N$=17 nodes and $L^e$=10 km, and $N$=21 nodes and $L^e$=7.5km. The in-between scenarios fall into a gray zone where neither Switched nor Relayed QKD is notably better.

Fig. 10 presents the absolute SKR consumption for uniform rings of $N$=9 and 21 nodes and adjacent node distances $L^e$= 1, 10 and 20 km. Note that absolute SKR consumption depends heavily on the specific SKR generation model used (theoretical based on PDS QKD modules), whereas their relative comparison (previous and next results) relies on the shape of the generation function and is more general. The performance of the Relayed architecture remains flat for all distances below $K$=25 km. On the other hand, we observe that increasing the distance affects the performance of the Switched architecture, as it uses longer links (chords) that surpass the flat SKR generation threshold ($K$=25 km). While for scenarios with small distances as this with $L^e$= 1, Switched outperforms, the Switched consumption rate ($C_S$) decreases as the distance increases, eventually falling below that of the Relayed architecture consumption rate ($C_R$).

We considered the above results as reference, as they were obtained using reference settings / realistic parameters: a maximum acceptable received power of $R_A$ = -6 dBm, and thus flat SKR generation up to $K$=25 km, a SKR generation performance loss $P$=2dB due to arbitrary QKD modules pairings and an optical penalty $O$=2dB due to the optical switches. To account for different acceptable received power (broader or narrower receiver dynamic range) and different pairing penalties (ranging from ideal to severe), we conducted additional comparisons, presented below.

As discussed in Section II, Switched QKD requires arbitrary pairing of QKD modules, but some modules are not designed for such operation. In the reference setting, a performance loss of $P$=2 dB was used to account for arbitrary modules matchings, considered to be realistic with current technologies. To study the effect of this parameter further, we examined the performance for an ideal case without performance losses ($P$=0) and highly incompatible modules with $P$=10dB, while keeping all other parameters the same as in the default configurations (flat SKR generation region $K$=25 km, $O$=2 dB, $R$=5 min, $T_{real}$=180 min). Relayed QKD is not affected by such incompatibility issues as key generation occurs only between adjacent node pairs and thus its performance did not change.

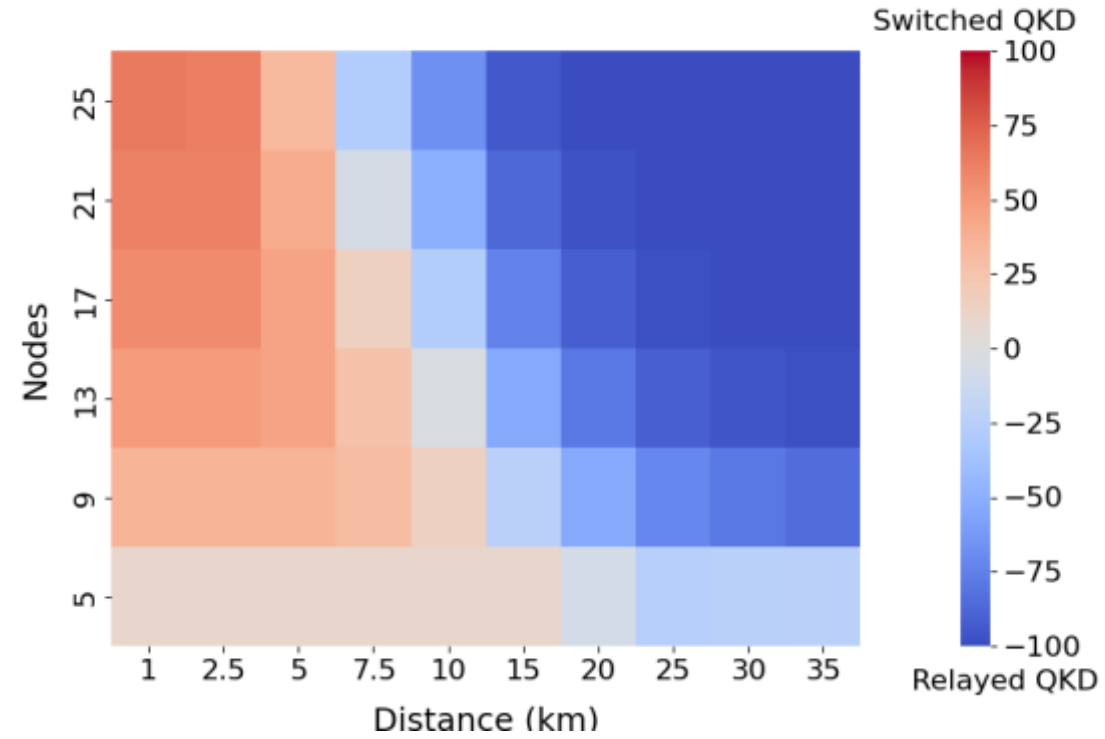

Fig. 9: Relative SKR consumption performance of Switched over Relayed architecture for varying nodes counts ($N$) and adjacent nodes distances ($L^e$), under reference parameters: flat SKR generation region $K$= 25 km, and for the Switched additional optical loss $O$=2 dB and SKR generation penalty $P$=2 dB.

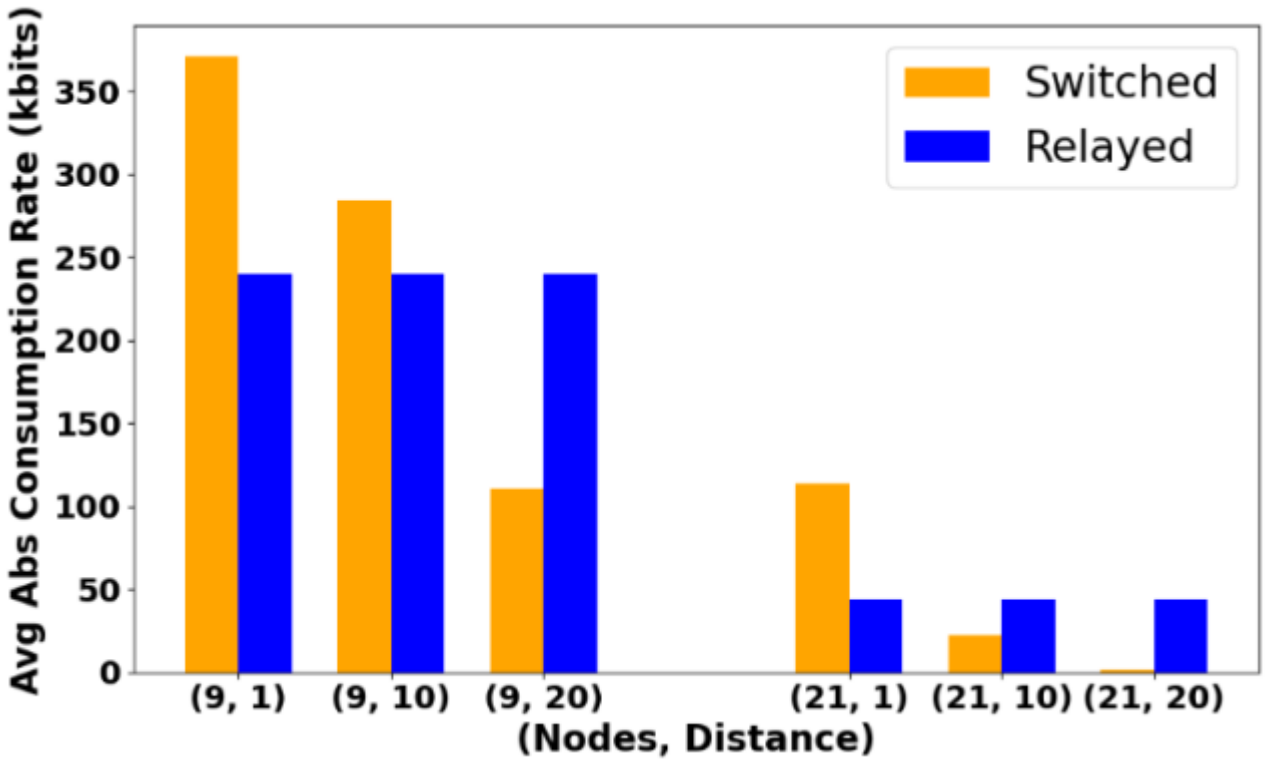

Fig. 10: Absolute SKR consumption performance of Switched ($C_S$) and Relayed ($C_R$) QKD architectures for $N$=9 and 21 nodes and adjacent nodes distance $L^e$=1, 10 and 20 km.

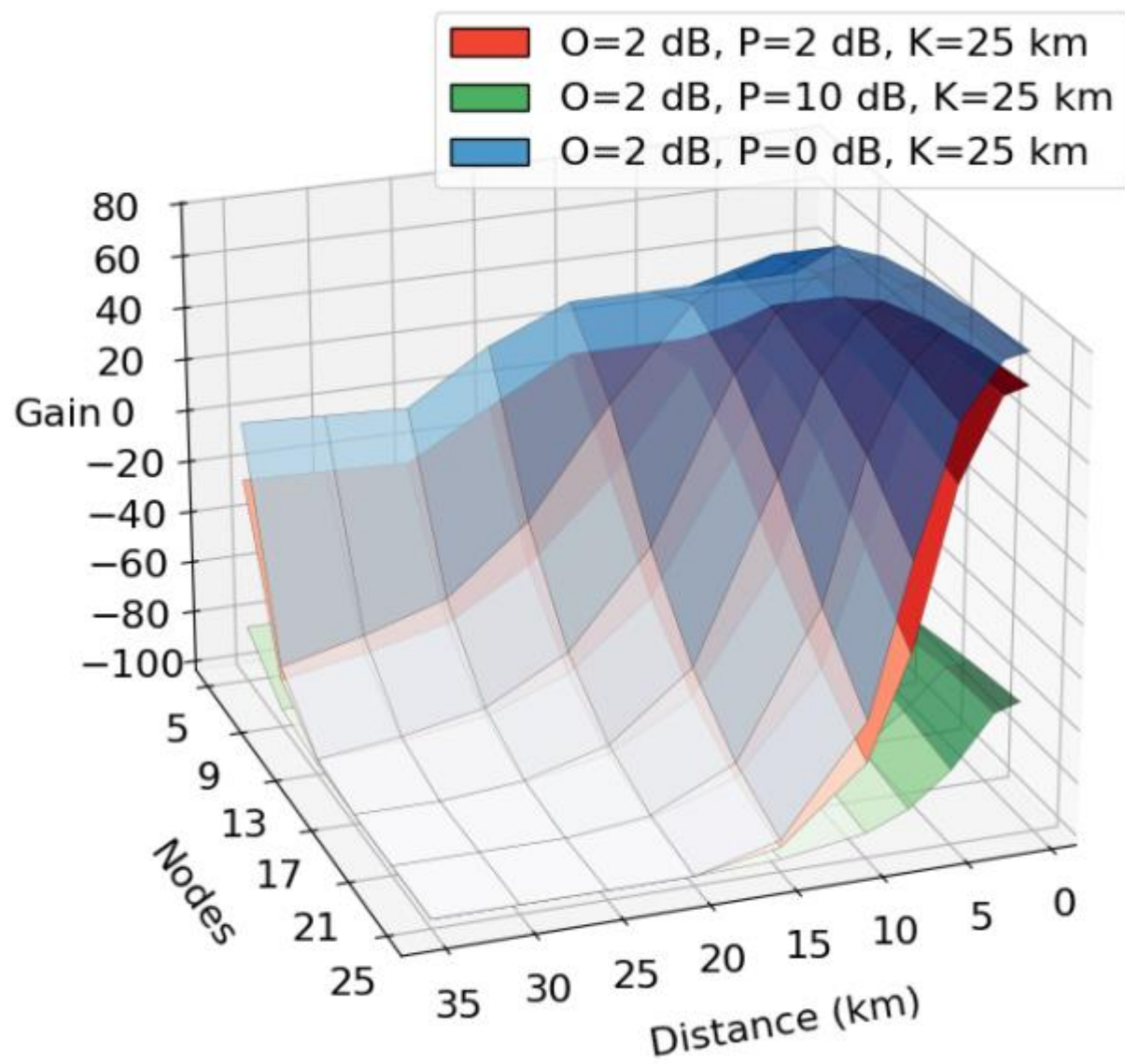

Fig. 11: Relative SKR performance of Switched over Relayed architecture, with additional SKR generation performance penalties of $P$=0, $P$=2 (reference), or $P$=10 dB in the Switched case due to arbitrary QKD modules pairings.

Fig. 11 shows the relative performance gain of the reference setting $P$= 2dB (red), ideal $P$=0 (blue) and worst-case/highly incompatible modules $P$=10dB (green). In the highly incompatible scenario ($P$=10dB), the performance of Switched QKD deteriorated significantly, making Relayed the only viable option. Conversely, in the ideal configuration ($P$ = 0), Switched QKD increased its relative gain to $f$ > 50 in scenarios that it was already better (25 < $f$ < 50) in the reference case, while surpassing $f$ > 25 in many neutral scenarios. Relayed, on the other hand, exhibited lower $f$ but remained clearly the best option for scenarios with $N$ > 9 and $L^e \geq 20$ km.

Summarizing, SKR generation deterioration of a few dB due to arbitrary pairing of QKD modules is acceptable, making the Switched architecture viable, and advantageous in dense networks with short link lengths and multiple hops. However, high incompatibility among QKD modules eliminates any benefit of Switched QKD.

Finally, we examined the impact of the dynamic receiver range. We simulated network scenarios with a maximum receiver power of $R_A$ = -12 dBm, corresponding to a $K$=50 km flat SKR generation region, and $R_A$ = 0 dBm, where SKR generation continuously increases as attenuation decreases, without a flat region ($K$=0).

Since SKR generation decreases exponentially with attenuation/distance, the Switched QKD architecture—using longer links (chords)—is more affected by long node distances. However, a longer flat SKR generation region (higher $K$) mitigates this effect: when chord lengths remain below the flat SKR generation threshold $K$, Switched QKD can generate keys even for distant nodes at rates equal to Relayed QKD between adjacent nodes. Thus, in scenarios with small $L^e$ , while both architectures generate keys at a comparable rate, Relayed QKD consumes a high percentage of its SKR generation to feed non-neighboring SAE pairs, making Switched QKD superior. A short SKR generation flat region (smaller $K$) benefits Relayed QKD at shorter $L^e$, since it increases the adjacent nodes SKR generation that would be flat otherwise.

Fig. 12 shows the relative performance for the reference setting (flat region $K$=25 km) and the two additional cases. With a $K$=50 km flat region (purple), Switched QKD—without increasing its peak performance—expands its advantage, outperforming Relayed QKD at node distances of $L^e$=15-20 km in smaller networks and 7.5-10 km in larger ones, and shifts more scenarios out of the neutral zone. In contrast, for configurations without a flat region $K$=0 (orange), Relayed QKD consistently outperforms Switched, except in dense networks with $L^e$=1 km nodes distance.

Summarizing, the maximum acceptable receiver power $R_A$ leads to flat SKR generation at low distances / attenuation. While this might seem like a limiting factor, it actually benefits Switched over the Relayed QKD. The reference setting of $R_A$=-6 dBm/ $K$=25km flat region which is typical in commercial QKD modules (e.g. Toshiba MU), suggests that such values — or even lower ones — result in substantial gains for the Switched QKD architecture in dense network scenarios.

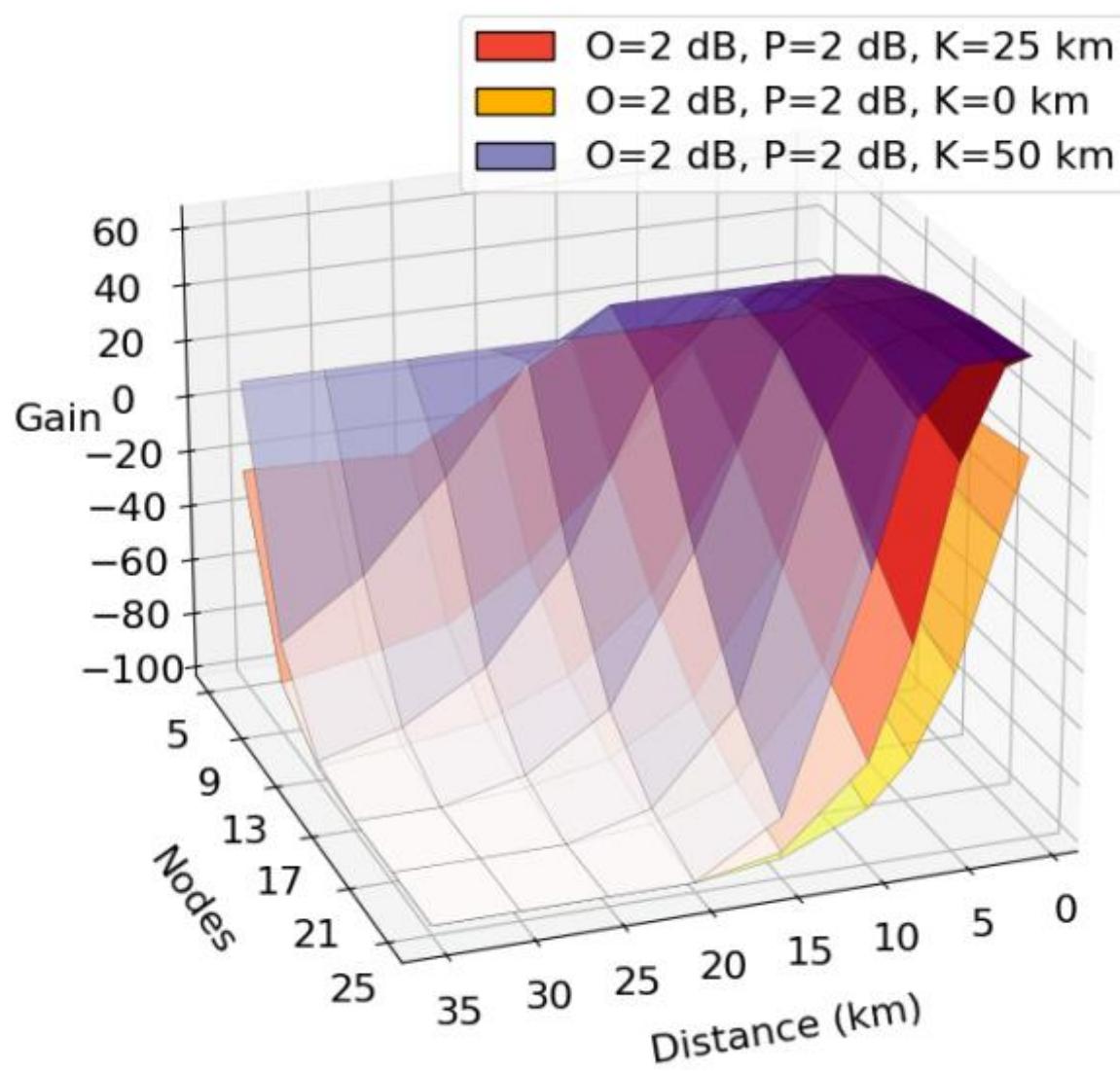

Fig. 12: Relative SKR performance of Switched over Relayed architecture with a SKR generation function *G* that exhibits a flat SKR generation up to *K*=0, 25 and 50 km.

## V. Conclusion

Relayed and Switched architectures are promising for network-wide QKD. Relayed QKD transmits keys over multi-link paths between quantum non-adjacent nodes, while switched QKD uses optical switches to dynamically connect arbitrary QKD modules, forming direct QKD links between nodes. A key advantage of switched QKD is that it distributes quantum keys end-to-end, whereas relayed QKD relies on trusted nodes. However, switched QKD requires arbitrary pairing of modules. We experimentally observed performance divergence in SKR generation when pairing unmatched QKD modules from three commercial vendors. To study this effect at network scale, we conducted a numerical analysis comparing Relayed and Switched QKD architectures, assuming uniform rings with equal adjacent node distances/attenuation and all-to-all key demands. We derived optimal configurations and analytical formulas for SKR consumption performance. Our results show that Switched QKD outperforms Relayed in dense networks (short distances and multiple hops). We also observed in the experiments that at short distances (i.e., low attenuation), the SKR generation remains flat to meet the maximum acceptable receiver power. While this flat SKR behavior might seem like a limiting factor, it benefits Switched over Relayed QKD— although Switched remains competitive even without this. However, the incompatibility of QKD modules and the observed SKR variations with unmatched pairings can substantially reduce Switched QKD's efficiency. To fully harness Switched architecture's benefits — the end-to-end quantum key distribution and low cost in dense networks — minimizing SKR penalties from unmatched pairings is crucial. A hybrid Switched and Relayed architecture could combine the benefit of both, enhancing network security and efficiency.

## Acknowledgments

This work was partially funded by the EU quantum flagship project QSNP (GA 101114043), HellasQCI project (GA 101091504), QRONOS (GA 16286) and Determined (GA 20501).